\documentstyle[12pt,graphicx]{article}
\newcommand{\be}{\begin{equation}}
\newcommand{\ee}{\end{equation}}
\newcommand{\bn}{\begin{eqnarray}}
\newcommand{\en}{\end{eqnarray}}
\begin{document}
\title{Nonrelativistic Quantum Analysis of the Charged Particle-Dyon System on a 
Conical Spacetime}
\author{A. L. Cavalcanti de Oliveira \thanks{E-mail: alo@fisica.ufpb.br} \ 
and E. R. Bezerra de Mello \thanks{E-mail: emello@fisica.ufpb.br}
\\
Departamento de F\'{\i}sica-CCEN\\
Universidade Federal da Para\'{\i}ba\\
58.059-970, J. Pessoa, PB\\
C. Postal 5.008\\
Brazil}
\maketitle
\begin{abstract}
In this paper we develop the nonrelativistic quantum analysis of the charged particle-dyon system in the spacetime produced by an idealized cosmic string. In order to do that, we assume that the dyon is superposed to the cosmic
string. Considering this peculiar configuration {\it conical} monopole harmonics are constructed, which are a generalizations of previous monopole harmonics obtained by Wu and Yang(1976 {\it Nucl. Phys. B} {\bf 107} 365) defined on a conical three-geometry. Bound and scattering wave functions are explicitly derived. As to bound states, we present the energy spectrum of the system, and analyze how the presence of the topological defect modifies obtained result. We also analyze this system admitting the presence of an extra isotropic harmonic potential acting on the particle. We show that the presence of this potential produces significant changes in the energy spectrum of the system.
\end{abstract}

\newpage
\renewcommand{\thesection}{\arabic{section}.}
\section{Introduction}

The analysis of the influence of a gravitational field and the topology of the spacetime on the quantum states of a physical system has received attention in last years. Along these lines of research the hydrogen atom, for example, has been analyzed in particular curved spaces \cite{AS,Parker,Parker1}. Parker \cite{Parker} 
showed that there appears shifts in the energy levels due to the local curvature, and this effect differs from the usual gravitational and Doppler shifts. These shifts would be appreciable only in region of strong gravitational field. Another line of investigation is to analyze the influence of the topology of the spacetime 
on a quantum system. Following this line of research, Marques and Bezerra have analyzed, under a nonrelativistic \cite{GB1} and relativistic \cite{GB2} point of views, the hydrogen atom placed on the spacetime produced by an idealized
cosmic string and global monopole specetimes. In their analysis they considered the hydrogen atom superposed to the defects. Although there is no local gravitational interaction between the atom and the defects, there appear 
modifications in the energy spectrum of the atom caused by the global aspects of the spacetime. Differently from the shifts caused by gravitational interaction, the shifts caused by the non trivial topology of the spacetime are global, i.e., they do not depend on the local value of curvature. Moreover in these calculations, the energy spectra are calculated exactly without any approximation.

Also recently Zhang {\it at al} \cite{Zhang1} and Zhang \cite{Zhang1} have analyzed the fermion-dyon system under a relativistic point of view. In these papers the authors have shown that the energy spectrum of this system is hydrogen-like, however with a delicate difference from the standard hydrogen-like one. Depending 
on the value of the parameter $q=eg$ the series of energy spectra do not match. The nonrelativistic quantum analysis of a charged particle-magnetic monopole system in the spacetime produced by a point-like global monopole has been developed
considering the presence of the electrostatic self-interaction acting on the particle \cite{Mello1}. There, bound and scattering states are investigated. As to bound states, we have shown that the non trivial topology of the spacetime 
modifies the energy spectrum of the system.

Cosmic strings \cite{Vilenkin1} and global monopole \cite{Vilenkin2} are exotic topological objects, which may be formed at phase transition of early Universe. Although none of these defects have been directly detected yet, cosmic strings have gained some recent interest since they are considered as a good candidate to explain some components of anisotropy on the cosmic microwaves background \cite{Sarangi}, gamma ray bursts \cite{Berezinski}, gravitational waves 
\cite{Damour} and highest energy cosmic rays \cite{Bhattacharjee}. Also they are thought to be important for the structure formation in the Universe due to their huge energy density per unit length \cite{Kible}. \footnote{ For cosmic strings formed at grand unified theory, the energy densities is of order than $10^{21}Kg/m$ 
and their radius $10^{-32}m$.}

One of the most attractive and elegant ideas of physics is the idea of {\it symmetry}. So, with the objective to obtain symmetry between electric and magnetic fields in the Maxwell equation, P. A. M. Dirac has proposed in 1931, 
the existence of a magnetic monopole \cite{Dirac}. In his formulation to the magnetic potential, the vector potential associated with this object is regular everywhere except to a semi-infinite line starting from the monopole. The magnetic
field produced by this potential is given by
\be
\vec{B}=g\frac{\hat{r}}{r^2} 
\ee
for points outside the singularity line.

In 1975, Wu and Yang presented the most and elegant formalism to describe Abelian pointlike magnetic monopole \cite{Wu-Yang}. In their formalism the vector potential is described by a singularity free expression. In order to
provide this formalism, they defined the vector potential $A_\mu$ in two overlapping regions, $R_a$ and $R_b$, which cover the whole space. Using spherical coordinate system, with the monopole at origin they have
chosen:
\begin{eqnarray}
\label{Ra}
R_a&:& 0 \leq \theta < \frac12\pi+\delta, \  r>0, \  0\leq \phi 
<2\pi \ , \\
R_b&:& \frac12\pi-\delta< \theta \leq \pi, \ r>0, \ 0\leq \phi < 2\pi ,\\ 
\label{Rab}
R_{ab}&:& \frac12\pi-\delta< \theta< \frac12\pi+\delta, \ r>0,  \
0\leq \phi < 2\pi \ , 
\end{eqnarray}
with $0<\delta \leq\frac12\pi$, and $R_{ab}$ being the overlapping region.

The only non vanishing components of vector potential are
\begin{eqnarray}
\label{A}
(A_\phi)_a&=&g(1-\cos\theta) \ ,
\nonumber\\
(A_\phi)_b&=&-g(1+\cos\theta) \ ,
\end{eqnarray}
$g$ being the monopole strength. In the overlapping region the non vanishing components are related by a gauge transformation
\begin{equation}
(A_\phi)_a=(A_\phi)_b+\frac ieS\partial_\phi S^{-1} \ ,
\end{equation}
where $S=e^{2iq\phi}$. In order to have single valued gauge transformation, we must have
\be
q=eg=n/2
\ee
in units $\hbar=c=1$. This is the Dirac quantization condition.

Schwinger \cite{Sch} and Zwanziger \cite{Zwa}, generalized the Dirac condition to allow for the possibility of the monopole to carry an electric charge. The quantum mechanical analysis of such two particles with electric and magnetic 
charges $(q_1,\ g_1)$ and $(q_2,\ g_2)$ provides the relation $q_1g_2-q_2g_1=n/2$. This quantization condition only determines the difference between the two electric 
charges.

Witten \cite{Witten} analysing the 't Hooft-Polyakov magnetic monopole system in the presence of a $CP-$violating term, showed that the electric charge of the dyon must be equal to $e\theta/2\pi$ plus an integer charge, being $\theta$ the
vacuum angle.

In this paper we shall study the nonrelativistic quantum motion of an electrically charged particle interacting with a dyon pierced by an idealized cosmic string. In order to develop this analysis we shall use the Schrodinger equation
\be
\label{S}
\left[-\frac1{2M}{\cal{\vec{D}}}^2+V\right]\Psi=i\partial_t\Psi \ ,
\ee
where
\be
{\cal{\vec{D}}}^2=\frac1{\sqrt{g^{(3)}}}{\cal{D}}_i\left(\sqrt{g^{(3)}}
g^{ij}{\cal{{D}}}_j\right)
\ee
being ${\cal{D}}_i=\partial_i-ieA_i$. $V$ is the Coulomb interaction, $M$ the mass of the particle and $g^{(3)}$ the determinant of the space components of the metric tensor.

Our main objective is to investigate how the presence of the cosmic string modifies some physical quantities such as, the energy spectrum of the system, and the phase shift for scattering states.

The analysis that will be developed here can be applied in high energy physics and in condensed matter physics as well. As will be shown in the section 3, linear defects in disordered solid or liquid crystal named disclinations can be treated, under a geometrical point of view, as having a metric structure similar to a cosmic string. So considering the possibility of a dyon be bound to a cosmic string or to a disclination, the system studied here may be relevant to provide informations about the quantum motion of a charged particle moving on the cosmos, and also for the motion of this particle in our neighbourhood.

The paper is organized as follows. In Sect. 2 we solve the Schrodinger equation (\ref{S}) considering stationary states, $\Psi(\vec{r},t)=e^{-iEt}\Psi_E(\vec{r})$, for the system under investigation. We present first the expression obtained for the monopole harmonics in a conical three-geometry, which we call by conical monopole harmonics. We also analyze the radial differential equation considering bound and scattering states. In Sec. 3 we extend the system admitting the presence of an extra isotropic harmonic potential acting on the particle. For this case we find exact solutions and energy spectrum for specific values of angular frequency $\omega$. Switching off the Coulomb interaction, the energy spectrum becomes much simpler. In this case we also analyze the modification in the energy spectrum due to the topology of the spacetime. Finally we leave for the Sec. 4 our conclusions and most relevant remarks.

\section{Nonrelativistic Quantum Mechanical Analysis of the System}

In this section we shall develop the nonrelativistic quantum analysis of a charged particle interacting with a dyon in the spacetime produced by an idealized cosmic string. In order to make this problem workable we shall consider the
dyon superposed to the linear defect. Moreover we shall adopt the singularity-free Wu-Yang formalism to describe the vector potential associated with the magnetic field. 

The spacetime produced by an idealized cosmic string, in spherical coordinate system reads
\be
\label{cs}
ds^2=-dt^2+dr^2+r^2d\theta^2+\alpha^2r^2\sin^2\theta d\phi^2 \ ,
\ee 
where $\alpha$ is a parameter smaller than unity which codifies the presence of the cosmic string. In fact $\alpha=1-4G\mu$, being $G$ the Newton gravitational constant and $\mu$ the linear mass density of the string. The spacetime given by (\ref{cs}) is locally flat and, consequently, there is no Newtonian potential.

According to previous analysis by Wu and Yang \cite{Wu-Yang1}, the solutions of the Schr\"odinger equation in the presence of such vector potential will not be ordinary functions, but instead, {\it section}, the solution assumes values $\Psi_a$ and $\Psi_b$ in $R_a$ and $R_b$, and satisfies the gauge transformation
\be
\Psi_a=S\Psi_b  \ 
\ee
in the overlapping region $R_{ab}$. 

The time-independent Schr\"odinger equation associated with this system is
\be
\left[-\frac1{2M}\frac1{\sqrt{g^{(3)}}}(\partial_i-ieA_i)\left[\sqrt{g^{(3)}}
g^{ij}(\partial_j-ieA_j)\right]+V(r)\right]\Psi(\vec r)=E\Psi(\vec r) \ .
\label{Sch}
\ee 

In order to solve the Schr\"{o}dinger equation above we shall adopt the usual approach: we write the solution in the form
\be
\label{Psi0}
\Psi(\vec r)=R(r)Y(\theta,\phi) \ .
\ee
Substituting (\ref{Psi0}) into (\ref{Sch}) and using the metric tensor given in (\ref{cs}), we obtain two different differential equations as shown below:
\bn
&&\frac1{\sin\theta}\partial_\theta[\sin\theta\partial_\theta Y(\theta,\phi)]+
\frac{1}{\alpha^2\sin^2\theta}
\partial^2_\phi Y(\theta,\phi)- \nonumber\\ 
&&\frac{2ieA_\phi}{\alpha^2\sin^2\theta}\partial_\phi Y(\theta,\phi)-
\frac{e^2A^2_\phi}{\alpha^2\sin^2\theta} Y(\theta,\phi)=-2M\lambda Y(\theta,\phi)
\label{ang}
\en
and
\be
\frac d{dr}\left(r^2\frac{dR(r)}{dr}\right)-2Mr^2\left(V(r)-E\right)R(r)=
2M\lambda R(r) \ , 
\label{rad}
\ee
being $\lambda$ a constant factor introduced to separate the original differential equation in two coupled ones. In order to procedure the analysis of the radial equation we must first solve (\ref{ang}). This analysis will be developed in the next subsection.

\subsection{Conical Monopoles Harmonics}

In order to obtain solutions for (\ref{ang}), we have to consider this equation in the regions $R_a$ and $R_b$ separately. Following the procedure adopted in \cite{Wu-Yang1} we can assume that
\be
Y(\theta,\phi)=\Theta(\theta)e^{i(m\pm q)\phi}\ ,
\label{Y}
\ee
where the positive (negative) sign in (\ref{Y}) refers to region $R_a$ ($R_b$). $m$ is the magnetic quantum number and $q=eg$.

Substituting (\ref{A}) and (\ref{Y}) into (\ref{ang}), and introducing a new variable $x=\cos\theta$, we obtain for both regions $R_a$ and $R_b$ the following differential equation
\be
(1-x^2)\frac{d^2\Theta(x)}{dx^2}-2x\frac{d\Theta(x)}{dx}-\frac{(m+qx)^2}
{\alpha^2(1-x^2)}\Theta(x)=-\bar{\lambda}\Theta(x)
\label{ang2}
\ee
with $\bar{\lambda}=2M\lambda$. In order to analyse the above equation we shall use the standard procedure: we shall investigate its behavior in the neighbourhood of its regular singular points $x=\pm 1$. After doing this development we can express the solution of (\ref{ang2}) by:
\be
\Theta(x)=(1-x)^{\frac{m+q}{2\alpha}}(1+x)^{\frac{m-q}{2\alpha}}G(x)
\label{sol1} 
\ee
where now $G(x)$ obeys the the following differential equation
\be
(1-x^2)\frac{d^2G(x)}{dx^2}-2[q_\alpha+(m_\alpha+1)x]\frac{dG(x)}{dx}-[m^2_\alpha
+m_\alpha-q^2_\alpha-\bar{\lambda}]G(x)=0 \ .
\label{sol2}
\ee
Defining a new variable $z=\frac{x+1}2$, it is possible to find solution for the equation adopting the series expansion 
\be
G(z)=\sum^\infty_{k=0}a_k z^k\ .
\label{serie}
\ee
Doing this we obtain the following recurrence relation
\be
a_{k+1}=\frac{(k+1+\beta+\sigma)k+\gamma}{(k+1+\beta)(k+1)}a_k\ ,
\ee
where
\be
m_\alpha=\frac{m}\alpha\ ,\ \ \ q_\alpha=\frac{q}\alpha\ ,\ \ \sigma=
m_\alpha+q_\alpha\ ,\ \ \ \beta=m_\alpha-q_\alpha\ ,
\ee
and
\be
\gamma=m^2_\alpha+m_\alpha-q^2_\alpha-\bar{\lambda}\ .
\ee

In order to have acceptable polynomials solutions, we must terminate the series at some finite value to $k$. Let us say $k=n$, with $n=0, \ 1, \ 2,\ ...$. In this case we get
\be
\bar{\lambda}=l_\alpha(l_\alpha+1)-q^2_\alpha
\label{auto}
\ee
with
\be
\label{l} 
l_\alpha=n+m_\alpha \ .
\label{l}
\ee
In the ordinary derivation of the associated Legendre polynomials, $P_n^m(x)$, the integer number $m$ was supposed to be a non-negative number. However, by using the Rodrigues' formula to define this function, this limitation can be relaxed.
In our development it was supposed that $m\pm q$ be also a non-negative number. By (\ref{l}) $l_\alpha$ is also a non-negative number. Because the number $n$ does not depend on the parameter $\alpha$, we can obtain a relation between the quantum numbers $l_\alpha$, and $l$ and $m$. This relation is 
\be
l_\alpha=l+|m|\left(\frac1\alpha-1\right) \ .
\ee

Substituting (\ref{auto}) into (\ref{sol2}), we obtain the Jacobi polynomials differential equation, whose solutions are \cite{G}: 
\be
P_n^{\sigma,\beta}(x)=\frac{(-1)^n}{2^nn!}(1-x)^{-\sigma}(1+x)^{-\beta}
\frac{d^n}{dx^n}\left[(1-x)^{\sigma+n}(1+x)^{\beta+n}\right] \ .
\ee
The solution of (\ref{ang}), in $R_a$, is
\be        
Y^{q_\alpha}_{l_\alpha,m_\alpha}(\theta,\phi)_a=N_{q_\alpha,l_\alpha}
(1-x)^{\frac{\sigma}{2}}(1+x)^{\frac{\beta}{2}}P_n^{\sigma,\beta}(x)
e^{i(m+q)\phi}\ ,
\label{Ya}
\ee
where
\be
N_{q_\alpha,l_\alpha}=\frac1{\sqrt{2\pi\alpha}}\left[\frac{(2n+\sigma+\beta+1)n!
\Gamma(n+\sigma+\beta+1)} {2^{\sigma+\beta+1}\Gamma(n+\sigma+1)\Gamma(n+\beta+1)}
\right]^{1/2}\ .
\ee
In $R_b$ the conical monopole harmonics have similar expressions changing $q$ by $-q$. 

So the conical monopole harmonics obey the eigen-value differential equation 
\be
{\vec L}^2_{q_\alpha}Y^{q_\alpha}_{l_\alpha,m_\alpha}=l_\alpha(l_\alpha+1)
Y^{q_\alpha}_{l_\alpha,m_\alpha} \ ,
\label{L2}
\ee
with
\bn
{\vec L}^2_{q_\alpha}&=&-\frac1{\sin\theta}\partial_\theta[\sin\theta
\partial_\theta]-\frac1{\alpha^2\sin^2\theta}\partial^2_\phi+\frac{2ieA_\phi}
{\alpha^2\sin^2\theta}\partial_\phi \nonumber \\
&+&\frac{e^2A^2_\phi}{\alpha^2\sin^2\theta}+q^2_\alpha \ .
\label{L22}
\en
Because $l_\alpha(l_\alpha+1)\geq q^2_\alpha$, we have $l_\alpha\geq\frac q{\alpha}$. In the case $\alpha=1$, Wu and Yang \cite{Wu-Yang1} showed that the allowed values for $l$ and $m$ are $l=|q|,\ |q|+1,\ |q|+2,\ ...$ and 
$m=-l,\ -l+1,\ ...\ l$.

\subsection{Radial Equation}

Now we are in position to analyze the radial differential equation (\ref{rad}). In order to do that we now, explicitly, use the electrostatic interaction between the charged particle and the dyon, given by the Coulomb potential
\be
V(r)=-\frac{eQ}r
\ee
where $-e$ is the charge of the particle and $Q$ the charge of the dyon. Substituting $R(r)=\frac{u(r)}r$, we get:
\be
\label{u1}
\frac{d^2u(r)}{dr^2}+\left[2ME+2M\frac{eQ}{r}-\frac{\bar{\lambda}_\alpha}{r^2}\right]
u(r)=0 \ ,
\ee
where
\be
\bar{\lambda}_\alpha=2M\lambda=l_\alpha(l_\alpha+1)-q^2_\alpha \ .  
\ee

Because it is our main interest to investigate how the presence of the linear defect modifies the energy spectrum of the charged particle-dyon system, and also the phase shift associated with scattering states, in the following we shall investigate (\ref{u1}) considering bound states, $E<0$, and scattering states, $E>0$.

\subsubsection{Bound States}

Considering negative energy states, we find the following solution
\bn
\label{u2}
u(r)&=&Ce^{-\kappa r}(2\kappa r)^{(1+\sqrt{1+4\bar{\lambda}_\alpha})/2}
\times\nonumber\\
&&{_1 F_1}\left(\frac12+\frac12\sqrt{1+4\bar{\lambda}_\alpha}-\gamma;1+
\sqrt{1+4\bar{\lambda}_\alpha};2\kappa r\right)\ ,
\en
where
\be
\kappa=\sqrt{-2ME}\ ,\ \ \ \ \ \ \ \ \ \gamma=\frac{MeQ}{\kappa}\ .
\ee

Because of the divergent behavior of the hypergeometric function, $_1F_1$, for large value of its argument, bound states solutions can only be obtained by imposing that this function becomes a polynomial of degree $N$. In this case 
(\ref{u2}) goes to zero at infinity. This condition is achieved by
\be
\frac12+\frac12\sqrt{1+4\bar{\lambda}_\alpha}-\gamma=-N \ \ \ N=0,\ 1,\ 2\ ...
\ee
Assuming this condition the energy spectrum is automatically provided:
\be
\label{En} 
E^q_{N,l,m}=-\frac{M(eQ)^2}2\left[N+\frac12+\frac12\sqrt{(2l_\alpha+1)^2-4q^2_\alpha}
\right]^{-2} \ .  
\ee        
    
From (\ref{En}), we can see that the energy spectrum associated with the system depend on the magnetic quantum number $m$. Consequently the presence of the cosmic string piercing the dyon reduces the degeneracy degree of the energy levels. The reason for this reduction is that the system now does not present spherical symmetry. In addition, because $l_\alpha\geq q_\alpha=q/\alpha$, we must have $l\geq q/\alpha$. So, for $q=1/2$, the lowest energy is attained by $N=0$, $l=3/2$ and $m=\pm 1/2$ \footnote{In fact for $\alpha\ll 1$, the allowed value to $l$ can be higher than $3/2$}. Consequently the degeneracy degree of this state is $2$. The introduction of an extra potential may remove this degeneracy. Before to consider this specific problem, we shall complete bellow, the analysis of finding the
complete eigen-function: The normalization constant of (\ref{u2}) is given by
\bn
|C^q_{N,l,m}|^2=\frac{2k\Gamma(N+\sqrt{(2l_\alpha+1)^2-4q_\alpha^2}+1)}
{\left[\Gamma(1+\sqrt{(2l_\alpha+1)^2-4q_\alpha^2})\right]^2(2N+\sqrt{(2l_\alpha
+1)^2-4q_\alpha^2}+1)N!} \ .
\en
In this way the solution associated with bound states reads
\bn
\Psi^q_{N,l,m}(\vec{r})&=&C^q_{N,\ l,\ m}\frac{e^{-\kappa r}}r(2\kappa r)^
{\frac12+\frac12\sqrt{1+4\bar{\lambda}_\alpha}}\times\nonumber\\
&&{_1 F_1}\left(-N;1+\sqrt{1+4\bar{\lambda}_\alpha}; 2\kappa r\right)
Y^{q_\alpha}_{l_\alpha,m_\alpha}(\theta,\phi) \ .
\label{Psi}
\en

Although the spacetime produced by an idealized cosmic string is locally flat \cite{Vilenkin3}, globally it is not. In fact the lack of global flatness is responsible for the existence of an induced repulsive electrostatic self-interaction
on a point charged particle \cite{Linet,Smith},  and also an induced electrostatic and magnetostatic self-interactions on long straight wires which present constant linear densities of charge and current parallel to
the topological defect \cite{Mello2}. As to a point charged particle, this repulsive self-interaction is given by
\be
U(\rho)=\frac{e^2L(\alpha)}{2\rho}=\frac{e^2L(\alpha)}{2r\sin\theta}\ ,
\label{U1}
\ee
where $L(\alpha)$ is a positive constant for $\alpha<\ 1$, given in terms of the
integral below \cite{Smith}
\be
L(\alpha)=\frac1{\alpha\pi}\int_0^\infty\frac{\coth(x/\alpha)-\alpha\coth(x)}
{\sinh(x)} \ .
\ee 

Developing the first order degenerate perturbation theory, the correction on the energy due to the perturbative potential is given by:
\bn
\left(\begin{array}{cc}
h_{11}&h_{12}\\
h_{21}&h_{22}
\end{array}
\right)
\left(\begin{array}{c}
c_1\\
c_2
\end{array}
\right)=\Delta E^{(1)}
\left(\begin{array}{c}
c_1\\
c_2
\end{array}
\right) \ ,
\en
where $h_{ij}=\langle\Psi^q_{\{i\}}|U(\rho)|\Psi^q_{\{j\}}\rangle$.

Below we shall calculate the energy correction to the ground state for the case $q=1/2$. Because of the azimuthal symmetry, the off-diagonal elements are both zero. So we get
\bn
h_{11}&=&\langle3/2,-1/2,1/2|U|3/2,-1/2,1/2\rangle
\label{35}
\en
and
\bn
h_{22}&=&\langle3/2,1/2,1/2|U|3/2,1/2,1/2\rangle \ .
\label{36}
\en

Substituting the wavefunctions correspondent to the states $\Psi^{1/2}_{0,3/2,\pm 1/2,1/2}\equiv|3/2,\pm1/2,1/2\rangle$ into (\ref{35}) and (\ref{36}), we obtain, after some intermediate steps for $\alpha=0.999$, the following results:
\bn
\Delta E^{(1)}_1=h_{11}=0.177Me^2Q^2L(0.999)
\en
and
\bn
\Delta E^{(1)}_2=h_{22}=0.161Me^2Q^2L(0.999) \ ,
\en
where $L(0.999)=3.9\times 10^{-4}$. 

As we can see the electrostatic self-interaction removes the degeneracy of the ground state associated with the system. Unfortunately, it is not possible to take in consideration the self-interaction in a complete form in the Hamiltonian
(\ref{Sch}). The only way to consider this potential in the system is through perturbation theory.

Before to finish this subsection we would like to call attention that our
expression obtained to the energy (\ref{En}) in the case $\alpha=1$, does 
not coincide with the nonrelatistic limit of the result given in \cite{Zhang1,Zhang2}
discarding the rest energy. However taking the limit $q=0$ this agreement is 
fullfiled.

\subsubsection{Scattering States}

Now let us consider positive energy states. In this case we find the following solution
\bn
u(r)&=&Ce^{ikr}(kr)^{(1+\sqrt{1+4\bar{\lambda}_\alpha})/2}\nonumber\\
&& {_1 F_1}\left(\frac1{2}+\frac1{2}\sqrt{1+4\bar{\lambda}_\alpha}-i\epsilon;1+
\sqrt{1+4\bar{\lambda}_\alpha};-2ikr\right)
\label{rad2}
\en
where
\bn
k=\sqrt{2ME}\,,\ \ \ \ \ \ \ \ \ \epsilon=\frac{MeQ}{k}\ .
\en
From the asymptotic behavior of the scattering wave-function, it is possible to obtain the phase shift, $\delta_l$, the most relevant parameter to calculate the scattering amplitude. The long distance behavior of (\ref{rad2}) reads 
\bn
\label{ua}
u(r)\approx C\cos\left[kr+\epsilon\ln(2kr)-\pi/4-\pi/2\left(\frac12
\sqrt{1+4\bar{\lambda}_\alpha}+\gamma_l\right)\right] \ .
\en

We can see that the phase shift presents two distinct contributions: \footnote{The logarthimic term in (\ref{ua}) is consequence of the long range Coulomb interaction. Because it does not depend on the partial wave analysis and varies slowly with the distance, it cannot be considered as a contribution for $\delta_l$ 
\cite{Schiff,Merz}}\\ 
$i)$ One from the Coulomb potential
\bn
\delta^{(1)}_l=\gamma_l=arg\Gamma\left(\frac12+\frac12\sqrt{1+4\bar{\lambda}_\alpha}
+i\epsilon\right) \ 
\en
which disappears if we take $Q=0$, and\\
$ii)$ the other from the modification on the effective angular quantum number, $\lambda_l$, due to the geometry of the manifold and also from the coupling between the electric charge of the particle with the magnetic filed of the dyon
\bn
\delta^{(2)}_l=\left(l+1-\frac{\sqrt{1+4\bar{\lambda}_\alpha}}2\right)\frac\pi4 \ .
\en

So the complete phase shift is given by the sum of both phases
\be
\delta_l=\delta^{(1)}_l+\delta^{(2)}_l \ .
\ee

\section{Harmonic Oscillator}

Linear topological defects are also present in condensed matter physics, they are vortices in superconductors or superfluid \cite{Kleiner}, and dislocations or disclinations in disordered solid or liquid crystal \cite{Kleman}. Cosmic strings and disclinations are linear defects that changes the topology of the the medium in a similar way. In fact the similarity between these two objects goes beyond the  topology: for some applications both kind of defects can be treated by the same geometrical methods \cite{Katanaev}. The linear defect disclination can be formed by either removing (positive disclination) or inserting (negative disclination) a wedge of material. Considering the defect infinitely long and along the $z-$axis, the effective three-dimensional metric tensor associated with the crystal has the same structure of a cosmic
string. In spherical coordinates the metric tensor can be given by the following line element
\be
dl^2=dr^2+r^2d\theta^2+\alpha^2r^2\sin^2\theta d\phi^2 \ .
\ee 
Being $\sigma$ the angle which defines the wedge, the parameter $\alpha$ is given by  $\alpha=1+\sigma/2\pi$ \cite{Katanaev}.

A few years ago Ruijgrok {\it at al} \cite{Ruijgrok} have analyzed the possibility of binding a magnetic monopole to an atomic nucleon or a to a molecule. In this paper the authors have found, using the Born-Oppenheimer approximation, an attractive effective monopole-atom potential and consequently proved the existence of a number bound states of Dirac monopole and hydrogen atom.  

The physical system that we want to investigate in this section is the charged particle-dyon one admitting existence of an extra isotropic harmonic potential acting on the particle. This system may be relevant to investigate the movement of 
a charged particle in a liquid crystal which presents a disclination piercing a dyon. The harmonic potential is introduced to provide bound states between the charged particle and the core of the dyon even for repulsive Coulomb interaction.

Considering this specific situation let us substitute the potential energy below
\be
V(r)=\frac12M\omega^2r^2-\frac{eQ}r \ ,
\ee
into (\ref{rad}). So the radial differential equation becomes:
\be
\label{Sch1}
\frac{d^2u(r)}{dr^2}+\left[2ME-M^2\omega^2r^2+2M\frac{eQ}r-\frac{\lambda_\alpha}
{r^2}\right]u(r)=0 \ ,
\ee
with
\be
\lambda_\alpha=l_\alpha(l_\alpha+1)-q_\alpha^2\ .
\ee

In order to analyse (\ref{Sch1}) let us first define a new dimensionless variable $\rho=\sqrt{M\omega}r$. Analyzing the behavior of this equation in the regions $\rho\approx0$ and $\rho\rightarrow\infty$, we can write for the solutions
the general form: 
\be
u(\rho)=\rho^{(1+\sqrt{1+4\lambda_\alpha})/2}e^{-\rho^2/2}F(\rho) \ ,
\ee
where $F(\rho)$ now satisfies the following equation
\be
\label{F}
\rho F''(\rho)+(\beta-2\rho^2)F'(\rho)+(\delta\rho-a)F(\rho)=0 \ ,
\ee
with
\bn
\beta=1+\sqrt{1+4\lambda_\alpha}\ ,\ \ \ \ \ \delta=b-2-\sqrt{1+4\lambda_\alpha}
\en
and
\bn
a=-2eQ\sqrt{\frac{M}{\omega}}\ , \ \ \ \ \ \ \ \ b=2\frac{E}{\omega} \ .
\en

The solution of the differential equation (\ref{F}) for $F(\rho)$ can be obtained as a power series \cite{Ver} 
\be
F(\rho)=\sum^\infty_{k=0}c_k\rho^k\ .
\ee
Substituting this series into (\ref{F}), we obtain the recurrence relations
\be
c_{k+2}=\frac{a}{(k+2)(k+1+\beta)}c_{k+1}+\frac{2k-\delta}{(k+2)(k+1+\beta)}c_k \ ,
\label{rel}
\ee
and
\be
c_1=\frac{a}{\beta}c_0\ .
\ee
The complete analysis of a differential equation similar to (\ref{Sch1}), has been develop by Vir\c cin \cite{Ver} and Vir\c cin {\it at al} \cite{Ver1}, investigating the problem of two anyons moving in a plane with Coulomb interaction, in the presence of an external uniform magnetic field. Moreover Furtado {\it  at al} \cite{Fur} analyzing the energy 
spectrum of an electron or hole in the presence of a magnetic field in the framework of the theory of defect of Katanaev and Volovich, also found a differential equation which presents a similar structure. Besides in \cite{Bezerra} we, analyzing the problem of two charged particle moving on a cone in the presence of a static uniform magnetic field plus a Coulomb interaction, also obtained a radial differential equation whose solution can be found using the procedure introduced by Ver\c cin. 

Let us now seek special exact solutions for (\ref{F}) which represent bound states. This is achieved by imposing that the series for $F(\rho)$ terminates at some finite value of $k$. Let $n$ be this maximum value. This condition implies that
\be
\delta=2n
\ee
and
\be
c_{n+1}=0 
\ee
simultaneously. Admitting that $c_0=1$ and using (\ref{rel}) we found:
\bn
c_2&=&\frac{a^2}{2\beta(\beta+1)}-\frac{\delta}{2(\beta+1)}\nonumber\\
c_3&=&\frac{a^3}{3!\beta(\beta+1)(\beta+2)}-\frac{a\delta}{3!(\beta+1)(\beta+2)}+
\frac{a(2-\delta)}{3!\beta(\beta+2)} \ .
\en

Let us now investigate the cases where $F_n(\rho)$ is a polynomial of degrees $n=1$ and $n=2$.
\noindent\\
$i)$ $n=1$. In this case we have 
\be
\delta=2 \ , 
\ee
and 
\be
F_1(\rho)=1+\frac{a}{\beta}\rho \ .
\ee
The energy eigenvalues are given by 
\be
E_{l,m}=\left(2+\frac{\sqrt{1+4\lambda_\alpha}}{2}\right)\omega\ .
\ee
The condition $c_2=0$ implies $a^2=\beta\delta=2\beta$. Consequently the angular frequency of the harmonic oscillator will be given by the equation
\be
\omega_{l,m}=\frac{2Me^2Q^2}{1+\sqrt{1+4\lambda_\alpha}}\ .
\label{om1}
\ee
Finally the energy associated with this state is
\be
E_{l,m}=e^2Q^2M\frac{4+\sqrt{1+4\lambda_\alpha}}{1+\sqrt{1+4\lambda_\alpha}} \ .
\ee
\\
\noindent\\
$ii)$ $n=2$. In this case we have
\be
\delta=4\ , 
\ee
and 
\be
F_2(\rho)=1+\frac{a}{\beta}\rho+\left(\frac{a^2}{\beta}-\delta\right)\frac{\rho^2}
{2(\beta+1)} \ .
\ee
The energy eigenvalues are given by
\be
E_{l,m}=\left(3+\frac{\sqrt{1+4\lambda_\alpha}}{2}\right)\omega\ .
\ee
The condition $c_3=0$ implies $a^2=4(2\beta+1)$. In this case the angular frequency of the oscillator is
\be
\omega_{l,m}=\frac{Me^2Q^2}{3+2\sqrt{1+4\lambda_\alpha}} \ .
\ee
The energy now becomes
\be
E_{l,m}=\frac{e^2Q^2M}2\frac{6+\sqrt{1+4\lambda_\alpha}}
{3+2\sqrt{1+4\lambda_\alpha}} \ .
\ee

From the results above we can see that: $i)$ The energy spectra associated with this system is composed by positive values. $ii)$ Since the product $eQ$ appears squared, the values of the angular frequency which correspond to exact energy eigen-values are the same. The difference shows up only in the wavefunction. $iii)$ There may exist infinite series solution for $F(\rho)$ with physically acceptable behavior \cite{Ver1}, however the possible values for the energy can only be obtained numerically. $iv)$ Finally we want to mention that as in the case studied in last section, the energies associated with the system depend on the quantum number $m$. 

Now, after this complete treatment, let us analyze the system considering the particular case where the Coulomb interaction is absent, i.e., $Q=0$. In this case the dyon becomes a magnetic monopole. The equation  of motion reads now
\be
\frac{d^2u(r)}{dr^2}+\left[2ME-M^2\omega^2r^2-\frac{\lambda_\alpha}{r^2}\right]u(r)
=0 \ .
\ee

The general regular solution is promptly obtained in terms of hypergeometric function:
\bn
R(r)&=&C_{l,m}e^{-\frac{M\omega}2r^2}r^{\frac12(-1+\sqrt{1+4\lambda_\alpha})}\times
\nonumber\\
&& {_1F_1}\left(\frac12+\frac{\sqrt{1+4\lambda_\alpha}}4-\frac b4,\ 1+
\frac{\sqrt{1+4\lambda_\alpha}}2;\ M\omega r^2\right) \ ,
\en
where $C_{l,m}$ is a normalization constant. In order to have bound states the hypergeometric function must becomes a polynomial of any finite degree $N$. This is achieved by imposing
\be
\frac12+\frac{\sqrt{1+4\lambda_\alpha}}4-\frac{b}4=-N \ .
\ee
This equation provides the quantization condition to the energy of the system which
reads
\be
E^q_{N,l,m}=\left(1+\frac{\sqrt{1+4\lambda_\alpha}}2+2N\right)\omega \ .
\ee

Taking $q=0$ and $\alpha=1$ in the above equation, we re obtain the energy spectrum associated with an isotropic three-dimensional harmonic oscillator. Finally we write down the complete solution to the bound wave function
\bn
\Psi_{l,m}^q(\vec{r})&=&C_{l,m}e^{-\frac{M\omega}2r^2}r^{\frac12(-1+
\sqrt{1+4\lambda_\alpha})}\times\nonumber\\
&&{_1F_1}\left(-N,\ 1+\frac{\sqrt{1+4\lambda_\alpha}}2;\ M\omega r^2\right)
Y_{l_\alpha,m_\alpha}^{q_\alpha}(\theta,\phi)\ .
\en

\section{Concluding Remarks}

In this paper we have analyzed the nonrelativistic quantum motion of a charged particle in the presence of a dyon superposed to an idealized cosmic string. Two specific physical situations have been investigated. The first one was to consider only the electromagnetic interaction between the charged particle with the dyon. In this case we could observe that the energy spectrum is similar to the hydrogen-like one. However, besides the dependence of the energy with the principal and angular quantum numbers, $N$ and $l$, respectively, there appears an extra dependence  with the magnetic quantum number $m$. This fact is a direct consequence of the presence of the cosmic string which reduces the degeneracy degree associated with a specific quantum level. Also the scattering wave-functions have been investigated. Analyzing its asymptotic behavior for large value of $kr$, we could observe that the phase shift presents two distinct contributions: one from the Coulomb interaction and the other from the modification on the effective angular quantum number.

The other problem analyzed was to consider the presence of an extra isotropic harmonic potential acting on the particle. The presence of this potential provides that all energy states are bound, even for repulsive Coulomb potential. In fact, as $r\to\infty$ both centrifugal and Coulomb potential of (\ref{Sch1}) tend to disappear while the harmonic potential grows without a limit. Also in this case the energy spectrum depends on the quantum number $m$. As a final application we
investigates the problem when the Coulomb potential is turned off. Once more the energy spectrum associated with this case is provided. 

Before to finish this paper we would like to make one more comment about the first system analyzed. The presence of the cosmic string modifies the hydrogen-like energy spectrum in a significative manner. The main reason is due the possible values that the ordinary angular quantum number can assumes. In the absence of cosmic strings, $l=|q|,\ |q|+1,\ |q|+2, \ ...$; however in its presence the minimum value for $l$ is bigger than $|q|$. By using the result (\ref{En}) for the energy spectrum, it is possible to see that for $q=1/2$, there is an increase of $68.6\%$ in the value of the energy for the lowest level considering $\alpha=0.999$, while for $q=1$ the increase is $66.4\%$. 

{\bf{Acknowledgments}}
\\       \\
We would like to thank Conselho Nacional de Desenvolvimento Cient\'\i fico e Tecnol\'ogico (CNPq.).

\end{document}